\documentclass[trackchanges]{aastex701}

\usepackage{amsmath} 

\newcommand{\dV}{\mathrm{d}V}
\newcommand{\dS}{\mathrm{d}S}

\usepackage{CJK}
\usepackage{amsmath,amssymb,amsthm,bm}
\usepackage{graphicx}   
\usepackage{color}      
\usepackage{soul,xcolor}
\usepackage{float}
\usepackage{caption}

\begin{document}
\begin{CJK*}{UTF8}{gbsn}

\title{Inertial-Range Energy Transfer Free from Isotropic Assumption in Turbulent Space Plasma}

\author{Zhuoran Gao (高卓然)}
\affiliation{Department of Physics and Astronomy,
University of Delaware, Newark, DE 19716, USA}
\email{veciam@udel.edu}
\author{Yan Yang (杨艳)}
\affiliation{Department of Physics and Astronomy,
University of Delaware, Newark, DE 19716, USA}
\email{yanyang@udel.edu}
\author{Francesco Pecora}
\affiliation{Department of Physics and Astronomy,
University of Delaware, Newark, DE 19716, USA}
\email{fpecora@udel.edu}
\author{Bin Jiang}
\affiliation{School of Mechanical Engineering and Mechanics, Xiangtan University, Xiangtan 411100, PR China}
\email{jiangbin@xtu.edu.cn}
\author{Kristopher G. Klein}
\affiliation{ Lunar and Planetary Laboratory, University of Arizona, Tucson, AZ, USA}
\email{kgklein@arizona.edu}
\author{Alexandros Chasapis}
\affiliation{Laboratory for Atmospheric and Space Physics, University of Colorado Boulder, Boulder, CO, USA}
\email{alexandros.chasapis@lasp.colorado.edu}
\author{Julia E. Stawarz}
\affiliation{School of Engineering, Physics, and Mathematics, Northumbria University, Newcastle upon Tyne, NE1 8ST, UK}
\email{julia.stawarz@northumbria.ac.uk}

\author{William H. Matthaeus}
\affiliation{Department of Physics and Astronomy,
University of Delaware, Newark, DE 19716, USA}
\email{whm@udel.edu}

\correspondingauthor{Yan Yang}
\email{yanyang@udel.edu}

\begin{abstract}
The idea of an energy cascade in the inertial range is often invoked in turbulent space plasmas to estimate the energy dissipation rate. Laws governing the behavior of third-order structure functions in the inertial range, so-called third-order laws, are among the few rigorous theoretical results quantifying cross-scale energy transfer. The widely used third-order-law derived rate assumes isotropy, which fundamentally conflicts with the anisotropic nature of space plasmas. Elementary questions persist regarding how such anisotropic energy cascades can be quantified using multi-spacecraft constellations. As the heliospheric community increasingly progresses towards multi-spacecraft, multi-scale
constellations, such as Plasma Observatory and HelioSwarm, we revisit these crucial issues pertinent to accurately measuring the inertial-range energy transfer. Here we make a systematic comparison between two methods: direction-averaging (DA) and lag polyhedral derivative ensemble (LPDE) to determine the full three-dimensional (3D) dependence of cross-scale energy transfer. We find that DA exhibits both polar and azimuthal dependence, but is insensitive to spacecraft configuration. By contrast, LPDE is strongly affected by spacecraft separation and tetrahedral shape, while being comparatively insensitive to the sampling trajectory. Our findings have direct implications for current and future multi-spacecraft missions. Both DA and LPDE will provide crucial information on the nature of turbulence in space and astrophysics.
\end{abstract}

\section{Introduction}\label{sec:Introduction}
Space plasmas are frequently collisionless or weakly collisional.
One of the most studied space plasmas is the solar wind \citep{BrunoCarbone13}, in which the mean-free path is of the order of 1 AU and collisions are too weak to establish a local equilibrium (Maxwellian distribution).
Energy dissipation for weakly collisional or collisionless plasma is of principal importance for addressing long-standing puzzles like the acceleration of energetic particles and the heating of the solar corona and solar wind. Collisionless dissipation has been investigated from different perspectives, including dissipation mechanisms \citep{Dmitruk04,MarkovskiiEA06,RetinoEA07,Howes2008kinetic,chandran2010perpendicular,klein2017diagnosing}, energy conversion channels \citep{Zenitani2011new,klein2017diagnosing,YangEA-PoP-17,YangEA-PRE-17,cassak2022pressure-I,cassak2022pressure-II,conleyKineticAnalogPressure2024}, and turbulence cascade \citep{TaylorKurienEyink2003,Sorriso07,MacBride08,Stawarz09,banerjee2016alternative,BandyopadhyayEA20-HallCasc,BandyopadhyayEA20-epsPSP,WuEA22-vKH}. Here we focus on energy cascade in magnetohydrodynamic (MHD) turbulence,  which has been adapted to quantify energy dissipation rate.

Turbulent flows are characterized by the disordered and chaotic behavior of the velocity and magnetic fields in space and in time. It is inherently a nonlinear phenomenon which couples motions at various scales. One of its defining processes is the cascade of energy (and other quantities conserved by the nonlinear interactions) across scales \citep{Taylor38,KarmanHowarth38,Kol41a,Kol41b}
wherein energy is transferred from large injection scales through an inertial range toward the dissipation range. A key quantity that characterizes this process is the energy cascade (or dissipation) rate, $\varepsilon$, which quantifies the flow of energy across different scales. 
The MHD turbulent cascade is much more complicated than its hydrodynamic counterpart. For example, the existence of two time-scales in MHD turbulence, one of the non-linear interaction and the other of the Alfv{\'e}n waves, makes it difficult to carry over the hydrodynamic turbulence theory in a straightforward manner. However, the analogous energy cascade process was still postulated in MHD turbulence, which was then supported by numerical simulations \citep{Muller00,MullerGrappin05,Aluie10} and spacecraft observations \citep{GoldsteinEA95}. 

While at kinetic scales, effects of charged particle motion must be considered. MHD remains a credible approximation for a kinetic plasma at scales large enough to be well separated from kinetic effects. This allows the inertial-range energy cascade process to be applied to weakly-collisional/collisionless plasmas, providing an estimate for the dissipation rate at the smallest scales without considering complex kinetic physics.

With the assumption of time stationary, spatial homogeinity, the existence of an inertial range, and finite dissipation rate, the original exact third-order formulation for isotropic, incompressible hydrodynamic turbulence \citep{KarmanHowarth38,MoninYaglom-vol2} has been extended to incompressible MHD \citep{Politano98a,Politano98b} and related models, making it possible to estimate $\varepsilon$ from field increments in both numerical simulations and in-situ plasma measurements \citep{TaylorKurienEyink2003,Sorriso07,MacBride08,Stawarz09,banerjee2016alternative}. 
The one-dimensional (1D, isotropic) form of the third-order law has been widely applied in observational studies. Under the isotropic assumption, $\varepsilon$ could be easily derived from field increments from single-spacecraft observations. 
It has been applied in the near-Earth solar wind \citep{Sorriso-ValvoEA07,MacBride08}, the Earth's magnetosheath from Magnetospheric Multiscale (MMS) data \citep{BandyopadhyayEA20-HallCasc}, the near-sun solar wind from Parker Solar Probe (PSP) data near perihelia \citep{BandyopadhyayEA20-epsPSP}, and the inner heliosphere from Helios 1 and Helios 2 \citep{WuEA22-vKH}. 
Although the 1D third-order law is notable in its simplicity and elegance, one should keep in mind the necessary assumption of isotropy to arrive at it.
The use of this form 
assumes that the mean flow direction is representative of the actual three-dimensional energy transfer. This inevitable limitation conflicts 
with the anisotropic nature of many space plasmas
\citep{HorburyEA12,OughtonEA15}.

It is important to recognize that the energy transfer flux varies systematically over the direction relative to the mean magnetic field \citep{VerdiniEA15,Schekochihin2022BiasedReview,JiangEtAl2023_JFM_hypervisc}. Therefore, keeping the full 3D information helps significantly in disentangling contributions from different directions to the energy transfer.
There have been attempts to refine the 1D form \citep{PodestaEA07-3rd,StawarzEA09,Galtier12,Coburn15} by making various 
assumptions about the structure and symmetry of the energy transfer flux, such as the axisymmetry along the mean magnetic field. 
Instead of introducing more assumptions, recent advances have been made to maintain the full 3D dependence by direction-averaging (DA) method \citep{TaylorEA03,wang2022strategies,JiangEtAl2023_JFM_hypervisc} and 3D derivative method--lag polyhedral derivative ensemble (LPDE) method \citep{PecoraEA23}, which properly account for the effect of anisotropy in the inertial range. The LPDE and DA methods have been proposed and applied to numerical simulations \citep{wang2022strategies,JiangEtAl2023_JFM_hypervisc,PecoraEA23}, while until recently their applications to in situ observations become possible \citep{Osman11-PRL,BandyopadhyayEA20-HallCasc,Pecora2023PRL}, due to the
necessary simultaneous multi-point measurements that span three-dimensional spatial directions.

Understanding of energy transfer in turbulent plasmas has greatly advanced due to the four-spacecraft tetrahedra missions such as the MMS \citep{BurchEtAl2016SSR} and Cluster \citep{Escoubet2001_Cluster}.
Cluster and MMS allowed us to probe plasmas along more spatial directions but only at single scales at any one point in time.
Therefore, in the majority of previous studies, we have to resort to Taylor's frozen-in hypothesis \citep{Taylor38} to evaluate energy transfer by the DA method, or obtain a partial energy transfer rate by the LPDE method \citep{Pecora2023PRL}. The community is now progressing towards a larger constellation of satellites to probe plasmas at multiple scales simultaneously, such as HelioSwarm \citep{Spence2019AGU_HelioSwarm_SH11B04,Klein2023}, consisting of 9 spacecraft, and Plasma Observatory \citep{retino2022particle,Marcucci2024}, consisting of 7 spacecraft. We can anticipate that significant advances in evaluating cross-scale energy transfer rates will become available with these and similar future missions.

In this work, we focus on the two representative approaches: (i) the direction-averaging (DA) third-order-law estimator \citep{TaylorEA03} and (ii) the lag polyhedral derivative ensemble (LPDE) method \citep{PecoraEA23}. Both methods avoid making an isotropic assumption, although they do so in different ways. In principle, both methods can provide precise cascade rates for arbitrary turbulence anisotropies. However, only lag vectors along a limited number of directions are available even in multi-spacecraft observations. The question is how
to make the best use of the limited number of lag vectors to implement LPDE
and DA methods in multi-spacecraft observations. To address this question, in this paper we present a detailed comparison between the DA and LPDE estimators to clarify their respective strengths, limitations, and regimes of validity in anisotropic 
MHD turbulence. We use a three-dimensional driven incompressible MHD simulation and construct virtual four-spacecraft measurements with controlled trajectory directions and tetrahedral baselines. This controlled setting allows a fair comparison of the two estimators under spacecraft-like sampling constraints, including limited angular coverage for DA and finite baseline/shape effects for LPDE.

The remainder of this paper is organized as follows. In Section~\ref{sec:Methodology}, we review the incompressible MHD framework, the third-order law, and the two estimators, DA and LPDE. We also describe the virtual-spacecraft setup used for the analysis. Section~\ref{sec:Results-DA} presents the results from the DA method, including its polar-angle and azimuthal-angle dependence. Section~\ref{sec:Results-LPDE} presents the results from the LPDE method, with emphasis on the effects of spacecraft separation, offset effect, trajectory independence, and threshold effect. Finally, Section~\ref{sec:Conclusion and Discussion} summarizes the main findings and possible refinements, and discusses their implications for future multi-spacecraft measurements.

\section{Theory and Methodology}\label{sec:Methodology}

In MHD turbulence, we use the Els\"asser variables $\boldsymbol z^\pm \equiv \boldsymbol v \pm \boldsymbol b$, where $\boldsymbol v$ is the velocity fluctuation and $\boldsymbol b$ is the magnetic fluctuation in Alfv\'en-speed units. For a spatial separation (lag) vector
$\boldsymbol\ell$, we define the corresponding increment
$\delta\boldsymbol z^\pm(\boldsymbol\ell)\equiv \boldsymbol z^\pm(\boldsymbol x+\boldsymbol\ell)
-\boldsymbol z^\pm(\boldsymbol x)$.
The MHD von K\'arm\'an--Howarth (vKH) equation \citep{Politano98a,Politano98b} is essential for the understanding of the energy cascade process, which reads
\begin{equation}
\frac{\partial}{\partial t}\left\langle \left|\delta\boldsymbol z^\pm\right|^2 \right\rangle
=
-\nabla_{\boldsymbol \ell}\!\cdot
\left\langle \delta\boldsymbol z^\mp\,\left|\delta\boldsymbol z^\pm\right|^2 \right\rangle
+ 2\nu \nabla_{\boldsymbol \ell}^2\left\langle \left|\delta\boldsymbol z^\pm\right|^2 \right\rangle
-4\,\varepsilon^\pm ,
\label{eq:VKH_MHD}
\end{equation}
where $\langle\cdot\rangle$ denotes an ensemble average, $\nabla_{\boldsymbol \ell}$ denotes the derivative in lag space, $\nu$ is the (kinematic) viscosity (here equal viscosity and resistivity are used), and ${\varepsilon^\pm = \nu \left\langle \left|\nabla \boldsymbol z^\pm\right|^2 \right\rangle }$ are the mean dissipation rates associated with the $\boldsymbol z^\pm$ energy. 
Note that because of the ensemble average and the assumption of homogeneity, all terms should be independent of any specific position in real space. The total mean dissipation rate is a constant for a simulation, and can be calculated as
\begin{equation}
\varepsilon_{\rm diss} = \frac{\varepsilon^+ + \varepsilon^-}{2}
= \nu \left\langle \boldsymbol{\omega}^2 + \boldsymbol{J}^2 \right\rangle ,
\label{eq:eps_total_def}
\end{equation}
where $\boldsymbol\omega\equiv\nabla\times\boldsymbol v$ is the vorticity and $\boldsymbol J\equiv\nabla\times\boldsymbol b$
is the electric current density.

Broadly, three scale regimes can be identified in MHD turbulence: a large-scale energy injection range, a small-scale dissipation range, and an intermediate inertial range connecting them.
In the statistically stationary limit and within the inertial range, the vKH equation reduces to the third-order law,
\begin{equation}
 \nabla_{\boldsymbol \ell}\!\cdot\left\langle \delta\boldsymbol z^\mp(\boldsymbol \ell)\,\big|\delta\boldsymbol z^\pm(\boldsymbol \ell)\big|^2 \right\rangle \equiv \nabla_{\boldsymbol \ell}\!\cdot\boldsymbol Y^\pm(\boldsymbol \ell)=-4\,\varepsilon^\pm,
\label{eq:PP_div_form}
\end{equation}
where
$\boldsymbol Y^\pm(\boldsymbol \ell)\equiv
\left\langle \delta\boldsymbol z^\mp(\boldsymbol \ell)\,\big|\delta\boldsymbol z^\pm(\boldsymbol \ell)\big|^2 \right\rangle$ is the Yaglom flux vector.
Note, under the assumption of homogeneity, this relationship holds in both isotropic and anisotropic cases.

Following \citet{Politano98a,Politano98b}, if isotropy  is assumed,
the Yaglom flux vector only has the radial component and  the divergence form in Eq.~\eqref{eq:PP_div_form} is equivalently reduced to the 1D form,
\begin{equation}
Y_\ell^{\pm}(\ell)
\equiv \left\langle \delta z_\ell^\mp(\boldsymbol{\ell})\,\big|\delta\boldsymbol z^\pm(\boldsymbol{\ell})\big|^2 \right\rangle
= -\frac{4}{3}\,\varepsilon^{\pm}\,\ell,
\label{eq:PP_4overd}
\end{equation}
where the longitudinal third-order structure functions
    $ Y_\ell^{\pm} =\langle \delta z_\ell^{\mp} |\delta\boldsymbol{z}^{\pm}|^2 \rangle$
are the projection of the Yaglom flux vectors along
  $\boldsymbol{\ell}$
  and 
  $\delta z_\ell^{\mp} = \delta\boldsymbol{z}^{\mp} \cdot \frac{\boldsymbol{\ell}}{\ell} $.
This law has been broadly used. However, isotropy could only be rarely assumed \citep{OsmanHorbury07,Verdini2015ApJ804119}, so it is necessary to take into account the directional dependence. Eq.~\eqref{eq:PP_div_form} has been implemented in several ways. In this work, we focus on two different methods: direction-averaging (DA) and lag polyhedral derivative ensemble (LPDE).

In the DA method, the integral form of Eq.~\eqref{eq:PP_div_form} is employed, and the data at different lag directions are collected along different trajectories in the simulation domain. In the analysis of space plasma observations, the Taylor frozen-in-flow hypothesis \citep{Taylor38} needs to be used to convert a time lag
$\tau$ into a spatial lag $\ell \simeq V_{\rm sw}\tau$ (where $V_{\rm sw}$ is the solar wind speed). 
In contrast, LPDE directly estimates the lag-space divergence of the Yaglom flux vector, as implemented in Eq.~\eqref{eq:PP_div_form}, which necessitates simultaneous multi-spacecraft measurements. In LPDE, the lag vectors are set by the
instantaneous inter-spacecraft baselines. In summary, DA and LPDE incorporate anisotropy in integral and differential forms of Eq.~\eqref{eq:PP_div_form}, respectively. 


\subsection{Direction-Averaging (DA) Method}
Eq.~\eqref{eq:PP_div_form} 
can be reformulated 
in terms of angle averaging. 
First, taking a volume integral over a sphere with radius $\ell = | \boldsymbol{\ell} |$ yields
\begin{equation}
  \iiint_{|\boldsymbol{\ell}|\le \ell} \nabla_{\boldsymbol \ell} \cdot \boldsymbol{Y}^{\pm} \, \dV
  = 
  \iiint_{|\boldsymbol{\ell}|\le \ell} -4 \varepsilon^{\pm} \, \dV
  =
  -\frac{16 \pi}{3} \varepsilon^{\pm} \ell^3
   .
\label{eq:div02_3rd_order_law}
\end{equation}
Using Gauss's theorem, this can be written as a surface integral,
\begin{equation}
   \oint_{|\boldsymbol{\ell}|=\ell} Y_\ell^{\pm} \, \dS 
 = 
   -\frac{16 \pi}{3} \varepsilon^{\pm} \ell^3, 
\label{eq:div03_3rd_order_law}
\end{equation}
In spherical coordinates, we express this 
in terms of the solid angle average of $Y^\pm_\ell$:
\begin{equation}
  \frac{1}{4\pi} \int_{0}^{2\pi} \int_{0}^{\pi} 
         Y_\ell^{\pm}  \sin\theta \, d\theta d\phi 
   = 
   - \frac{4}{3} \varepsilon^{\pm} \ell, 
 \label{eq:3d_3rd_order_law}
\end{equation}
where $\theta$ represents the polar angle (from the $\boldsymbol{B}_0$ axis) and $\phi$ the azimuthal angle. 
Considering that no assumptions about rotational symmetry are made in going from 
Eq.~\eqref{eq:PP_div_form} to 
Eq.~\eqref{eq:3d_3rd_order_law}, the physical content of 
Eq.~\eqref{eq:3d_3rd_order_law} is as general as the derivative form Eq.~\eqref{eq:PP_div_form}. 
The full generality of 
Eq.~\eqref{eq:3d_3rd_order_law} follows from the rigorous theorem given by \cite{NieTanveer99} and restated in more accessible terms by \cite{TaylorEA03,wang2022strategies,JiangEtAl2023_JFM_hypervisc}. However, Eq.~\eqref{eq:3d_3rd_order_law} is simpler in the sense that accurate determination of integration only requires the longitudinal component of the Yaglom flux vectors, $Y_\ell^{\pm}$, 
on the spherical surface spanned by the coordinates $(\theta, \phi)$ in the 3D lag space.   

In the DA method, the solid angle average in Eq.~\eqref{eq:3d_3rd_order_law} is realized via 
\begin{equation}
\widetilde{Y}_{\ell}^{\pm}
=
\frac{1}{4\pi}\int_{0}^{2\pi}\int_{0}^{\pi}
Y_{\ell}^{\pm}(\theta,\phi)\,\sin\theta\,\mathrm{d}\theta\,\mathrm{d}\phi
\approx
\frac{\displaystyle\sum_{j=1}^{N_{\phi}}\sum_{i=1}^{N_{\theta}}
Y_{\ell}^{\pm}(\theta_i,\phi_j)\,\sin\theta_i}
{\displaystyle N_{\theta}\sum_{i=1}^{N_{\theta}}\sin\theta_i}\, .
\label{eq:discre_Y}
\end{equation}
We discretize the sphere by choosing
$N_{\theta}$ polar angles $\{\theta_i\}$ and $N_{\phi}$ azimuthal angles $\{\phi_j\}$, so that the total number of sampled directions is $N_\Omega=N_{\theta}N_{\phi}$. The factor $\sin\theta_i$ comes from the solid-angle element $\mathrm{d}\Omega=\sin\theta\,\mathrm{d}\theta\,\mathrm{d}\phi$ and acts as the weight for each ring with constant $\theta_i$ and full coverage of $\phi$.

\subsection{Lag Polyhedral Derivative Ensemble (LPDE) Method}
\label{sec:LPDE}
The direction-averaging method is realizing Eq.\eqref{eq:3d_3rd_order_law} (the integral form of Eq.~\eqref{eq:PP_div_form}), while the recently proposed lag polyhedral derivative ensemble (LPDE) method makes use of the differential form as Eq.~\eqref{eq:PP_div_form}. 
Even though it is trivial to calculate the differential form as Eq.~\eqref{eq:PP_div_form} in the analysis of numerical simulation data, one could encounter challenges in the absence of adequate sample directions, which is typical in space plasma observations. Therefore, the calculation of $\nabla_{\boldsymbol{\ell}}\!\cdot\mathbf{Y}^{\pm}$ is infeasible prior to the advent of multi-spacecraft constellations, such as MMS and Cluster. 
With at least four spacecraft, \cite{PecoraEA23} proposed the lag polyhedral derivative ensemble (LPDE) method, which is actually a technique to compute gradients in lag space, i.e., $\nabla_{\boldsymbol{\ell}}$.

Here, we briefly review the procedures of LPDE and one can refer to \cite{PecoraEA23,Pecora2023PRL} for more details. The first step is the selection of tetrahedra in lag space. From the offset of 4 spacecraft in space, 6 baselines in total are formed by pairs of spacecraft, each of which can serve as a vector in lag space.
By including the additional 6 lag vectors along the opposite directions of the 6 baselines, we can have 12 lag vectors.
From the lag vectors, candidate tetrahedra are constructed by selecting 4 lag vectors from 12 lags. Thus, we have ${12\choose4}=495$ candidate lag-tetrahedra (tetrahedra in lag space) in total. We first remove the lag-tetrahedra with 0 volume, which are constructed by two lags and their opposite directed pair, i.e., $\{\mathbf{a},-\mathbf{a},\mathbf{b},-\mathbf{b}\}$, leaving us 480 lag-tetrahedra. To avoid double counting, each lag-tetrahedron is paired with its opposite, and we only keep a single representative of each pair. The remaining 240 lag-tetrahedra are used for the subsequent analysis. In principle, one can compute $\nabla_{\boldsymbol{\ell}}$ for each lag-tetrahedron via the curlometer technique we will discuss later. However, as suggested in \citet[Sec.~13.3.3, p.~328]{RobertEtAl1998_Tetrahedron}, the lag-tetrahedron
shape and size could significantly impact the numerical error of the calculation of $\nabla_{\boldsymbol{\ell}}$.  We can characterize the tetrahedron
shape in real and lag spaces using the elongation $E$ and planarity $P$ (both constructed from the eigenvalues of the
volumetric tensor) and define the combined quality metric
$d_{EP}=\sqrt{E^{2}+P^{2}}$.
Only lag-tetrahedra satisfying $d_{EP}\le 0.85$ will be retained, which
removes highly elongated or nearly planar lag-tetrahedra while keeping a well-conditioned
ensemble for the subsequent calculation.

As an additional condition, here we are computing the third-order law as in Eq.~\eqref{eq:PP_div_form}, which is valid at the inertial range. Therefore, we impose an extra threshold to remove lag-tetrahedra that are too small to plausibly lie in the inertial range. In practice, we will only select lag-tetrahedra, in which $\ell_{\mathrm{meso}} \ge 1\,\ell_K$ ($\ell_{\mathrm{meso}}$ is the distance from the origin to the lag-tetrahedral mesocenter in lag space and 
$\ell_K=(\nu^3/\varepsilon)^{1/4}$ is the Kolmogorov scale of the particular field). This prevents the estimate from being contaminated by lag-tetrahedra that are unlikely to lie in the inertial range.

The second step is the calculation of $\nabla_{\boldsymbol{\ell}}$ using the curlometer technique \citep{PaschmannDaly2008MSAMR}.
For each retained lag-tetrahedron, we first consider a generic vector field $\boldsymbol F(\boldsymbol x)$ sampled at the
four vertices $\boldsymbol x_\alpha$ ($\alpha=1,2,3,4$). 
Its divergence can be approximated by the standard reciprocal-vector formula \citep{PaschmannDaly2008MSAMR}
\begin{equation}
\nabla_{\boldsymbol x}\!\cdot \boldsymbol F
\;\approx\;
\sum_{\alpha=1}^{4} \boldsymbol F(\boldsymbol x_\alpha)\cdot \boldsymbol k_\alpha ,
\label{eq:tet_divF}
\end{equation}
where $\{\boldsymbol k_\alpha\}$ are the reciprocal vectors (the reciprocal basis of the lag-tetrahedron), defined as
\begin{equation}
\boldsymbol k_\alpha
=
\frac{\boldsymbol x_{\beta\gamma}\times \boldsymbol x_{\beta\lambda}}
{\boldsymbol x_{\beta\alpha}\cdot\left(\boldsymbol x_{\beta\gamma}\times \boldsymbol x_{\beta\lambda}\right)},
\qquad
\boldsymbol x_{\alpha\beta}=\boldsymbol x_\beta-\boldsymbol x_\alpha,
\label{eq:tet_recip_k}
\end{equation}
and $(\alpha,\beta,\gamma,\lambda)$ is a cyclic permutation of $(1,2,3,4)$.

In our application, the ``vertices'' are points in lag space, i.e., $\boldsymbol x\equiv\boldsymbol \ell$,
and the vector field is the third-order (Yaglom) flux vector $\boldsymbol F\equiv\boldsymbol Y^\pm$. Therefore,
\begin{equation}
\nabla_{\boldsymbol \ell}\!\cdot \boldsymbol Y^\pm(\boldsymbol \ell)
\;\approx\;
\sum_{\alpha=1}^{4} \boldsymbol Y^\pm(\boldsymbol \ell_\alpha)\cdot \boldsymbol k_\alpha ,
\label{eq:lpde_divY}
\end{equation}
where the reciprocal vectors $\boldsymbol k_\alpha$ are constructed from the lag-tetrahedron vertices $\{\boldsymbol \ell_\alpha\}$.
From Eq.\eqref{eq:PP_div_form}, the cascade rate is then estimated as
\begin{equation}
\varepsilon^{\pm}
\;=\;
-\frac{1}{4}\,\nabla_{\boldsymbol \ell}\!\cdot \boldsymbol Y^{\pm}
\;\approx\;
-\frac{1}{4}\sum_{\alpha=1}^{4}\boldsymbol Y^{\pm}(\boldsymbol \ell_\alpha)\cdot \boldsymbol k_\alpha.
\label{eq:eps_lpde}
\end{equation}

At the end, we will be able to obtain a cascade rate estimate from each lag-tetrahedron at the scale $\ell_{\mathrm{meso}}$ determined by the mesocenter of the lag-tetrahedron. This procedure has been applied to MMS data in \cite{Pecora2024arXiv2410.16099,Pecora2023PRL}.

\subsection{Virtual Spacecraft Analysis}
\label{sec:virtual}
To assess the performance of the DA and LPDE methods, we analyze a three-dimensional driven incompressible MHD simulation with mean magnetic field $B_0$ from \citet{JiangEtAl2023_JFM_hypervisc} and
construct virtual four-spacecraft measurements with controlled trajectory directions and tetrahedral baselines.
That is, we employ four spacecraft in tetrahedral configurations to fly through numerically generated turbulent fields, mimicking satellite (e.g, MMS and Cluster missions) flights through solar wind and magnetosheath turbulence.

\begin{figure}[H]
  \centering
  \includegraphics[width=0.5\textwidth]{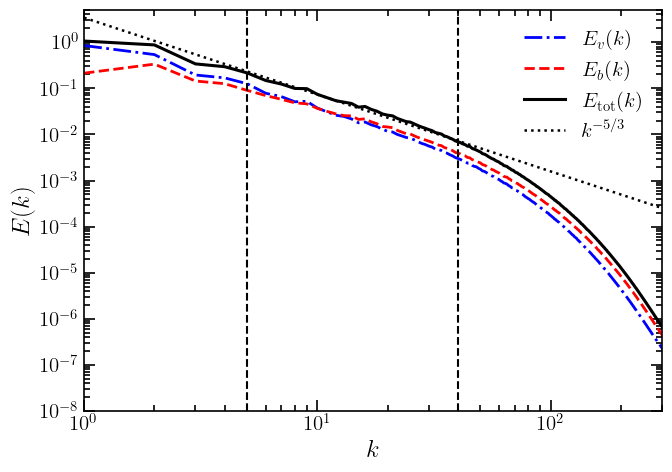}
  \caption{Energy spectrum of the simulation. The blue dash-dotted and red dashed curves show the kinetic and magnetic energy spectra, respectively, while the black curve shows the total energy spectrum. The black dashed line indicates a reference $-5/3$ power law. The vertical dashed lines mark the inertial range.}
  \label{fig:energy_spectrum}
\end{figure}

The simulation is carried out with $1024^3$ grid points in Fourier space using a pseudo-spectral method. The two-thirds rule is used for dealiasing and the second-order Adams--Bashforth method is used for time integration \citep{Orszag1971,Orszag1972,gottlieb1977numerical,Orszag_Tang_1979,canuto2007spectral}. The simulation domain is a three-dimensional periodic box with a size of $2\pi$ in every direction, and each grid cell $\Delta$ corresponds
to a physical length of $\Delta=0.0061$. 
An external mechanical force is applied to the first two wavenumber shells ($k=1, 2$) of the velocity field \citep{yang2021effects} to achieve a statistically stationary state.  An external mean magnetic field, $\mathbf{B}_0= 2 \hat{\mathbf{z}}$, is imposed along the $z$-direction. 
The simulation is initialized with random velocity and
magnetic fluctuations within the wavenumber band $k \in [1,5]$, which follow the prescribed spectrum $1/\left[1 + (k/k_0)\right]^{11/3}$ and $k_0 = 3$. The initial kinetic and magnetic energies are equal, i.e.\ $E_v = E_b = 0.5$. The cross-helicity is negligible. Equal viscosity and resistivity $\nu = \eta = 8 \times 10^{-4}$ are used for this simulation. After $\sim5$ large-eddy turnover times, the simulation reaches a statistically stationary state and we perform our analysis at this instant of time. By using Eq.~\eqref{eq:eps_total_def}, $\varepsilon_{\rm diss}=1.66$ is derived.
Figure~\ref{fig:energy_spectrum} shows the kinetic and magnetic energy spectra of the simulation. The approximate inertial range, identified with a $k^{-5/3}$ scaling, is marked in the plot, corresponding to \(5<k<50\).

We create virtual trajectories within the simulation domain to mimic satellites orbiting through real space plasma turbulence and collect virtual spacecraft time series. Note that, we are only using one time snapshot of the simulation, and the turbulence does not evolve in time as the virtual spacecraft moves through it. Here we are using four spacecraft, like the MMS and Cluster missions. Since the time evolution of the real space plasma turbulence is much faster than the spacecraft drifting timescale, the inter-spacecraft separations along each trajectory are fixed. So the trajectories of the virtual spacecraft are parallel straight lines, with specified polar and azimuthal angles. 
With $\mathbf{B}_0$ aligned with $\hat{\mathbf{z}}$, each straight-line sampling
direction is parameterized by a polar angle $\theta$ (relative to $\hat{\mathbf{z}}$) and an azimuthal angle
$\phi$ in the perpendicular plane.

Due to the periodic boundary conditions of the simulation, as the virtual spacecraft moves, the trajectories periodically cross the simulation box several times as shown in Figure 2(a) in \cite{PecoraEA23}.
One might also note that the length of the wrapped trajectory could be very short along certain directions, such as $\theta=0^\circ, 45^\circ, 90^\circ$ or $\theta$ near multiples of $\pi/2$.
So we avoid directions aligned with the coordinate axes or close to the principal symmetry planes of the cube, and instead select a set of directions that is approximately uniform on the sphere while being offset from these directions. Specifically, we use 42 directions arranged in seven polar angles
and six azimuthal angles,
\[
\theta \in \{12^\circ,\,24^\circ,\,36^\circ,\,48^\circ,\,60^\circ,\,72^\circ,\,84^\circ\},\qquad
\phi \in \{0^\circ,\,60^\circ,\,120^\circ,\,180^\circ,\,240^\circ,\,300^\circ\}.
\]
Along each trajectory, the sampling step is about $1.7\Delta$. 
To reduce potential dependence on the initial position, we assign distinct starting points for all 42 trajectories. In practice, the resulting sample lengths vary across directions, spanning from $2.5\times10^{5}$ to $4.7\times10^{6}$ data points. Since the sampling step we are using here is very close to the resolution of the simulation, we can expect that finer sampling steps with more data points along the trajectory will not significantly affect the results. 

Finally, we need to determine the tetrahedral formation of the four virtual spacecraft. We place four virtual spacecraft at fixed offsets forming an irregular tetrahedron with an non-regularity level of $\sigma$, which is a dimensionless scalar controlling the deformation from the regular tetrahedron.
The transformation matrix is $\mathbf{A}=\mathbf{I}+\sigma\,\mathbf{M}$, where $\mathbf{I}$ is the $3\times3$ identity matrix, $\mathbf{M} \equiv
\begin{pmatrix}
0.30 & 0.05 & -0.02 \\
-0.04 & -0.20 & 0.06 \\
0.03 & 0.02 & 0.10
\end{pmatrix}$ is a fixed $3\times3$ perturbation matrix that sets the deformation pattern,
and varying $\sigma$ only changes the deformation strength. After this deterministic deformation is applied, we further add a small random perturbation to the vertex coordinates in order to introduce additional irregularity into the tetrahedral shape. In practice, each Cartesian coordinate of the vertex is given an independent Gaussian-distributed random offset with zero mean. The amplitude of this additional random perturbation is set by $0.1\sigma$. Here we are using a fixed $\sigma=0.5$. The dependence on $\sigma$ will be discussed in later sections. 

Meanwhile, the interspacecraft separation also needs to be set up. Here six different offsets with the shortest edge $\{10,\,20,\,30,\,40,\,50,\,60\}\,\ell_K$ are used. In this simulation, $\ell_K$ is comparable to the grid spacing, with $\ell_K \approx 0.68\,\Delta$. The inter-spacecraft separations of $L_{\rm inter}=\{\,50,\,60\}\,\ell_K$ are in the inertial range and $L_{\rm inter}=\{10,\,20,\,30,\,40\}\,\ell_K$ are roughly in the dissipation range.

\section{Results of Direction-Averaging (DA) Method}\label{sec:Results-DA}
After sampling the fields along each virtual-spacecraft trajectory, we compute the longitudinal third-order structure
functions $Y_\ell^\pm(\boldsymbol{\ell})=\langle \delta z_\ell^\mp\,|\delta\boldsymbol z^\pm|^2\rangle$. 
Note that the most general form of $Y_\ell^\pm(\boldsymbol{\ell})$ should be a function of $\ell$, $\theta$ and $\phi$, i.e., $Y_\ell^\pm(\ell, \theta, \phi)$\citep{JiangEtAl2023_JFM_hypervisc}. In the isotropic case, it is reasonable to assume that $Y_\ell^\pm(\ell, \theta, \phi)$ is statistically independent of $\theta$ and $\phi$, i.e., $Y_\ell^\pm(\ell)$ shown in Eq.~\eqref{eq:PP_4overd}. In the anisotropic case, the turbulence has often been assumed to be statistically axisymmetric about the direction of the mean magnetic field (also called azimuthally symmetric or cylindrically symmetric) \citep{StawarzEA09,Coburn15,Galtier12,PodestaEA07-3rd}, i.e., $Y_\ell^\pm(\ell, \theta)$. Anisotropic energy transfer and its dependence on directions have been investigated in several MHD simulation studies \citep{VerdiniEA15,JiangEtAl2023_JFM_hypervisc,wang2022strategies}, which support that full 3D information should be incorporated, and the most general form $Y_\ell^\pm(\ell, \theta, \phi)$ should be explored.

Before performing the
solid-angle average in the DA procedure (Eq.~\eqref{eq:3d_3rd_order_law} and its discrete form in
Eq.~\eqref{eq:discre_Y}), it is useful to define a direction-by-direction, scale-dependent third-order estimate along each sampled lag direction
\begin{equation}
\mathcal{E}_3(\boldsymbol{\ell})\equiv -\frac{3}{8}\,\frac{Y_\ell^+(\boldsymbol{\ell})+Y_\ell^-(\boldsymbol{\ell})}{\ell},
\label{eq:E3_def}
\end{equation} 
In this section, we investigate the angular dependence of the energy transfer quantified by Eq.~\eqref{eq:E3_def}
and test the feasibility of the DA method.   

\subsection{Polar angle ($\theta$) dependence}
\begin{figure}[H]
  \centering
  \includegraphics[width=0.48\textwidth]{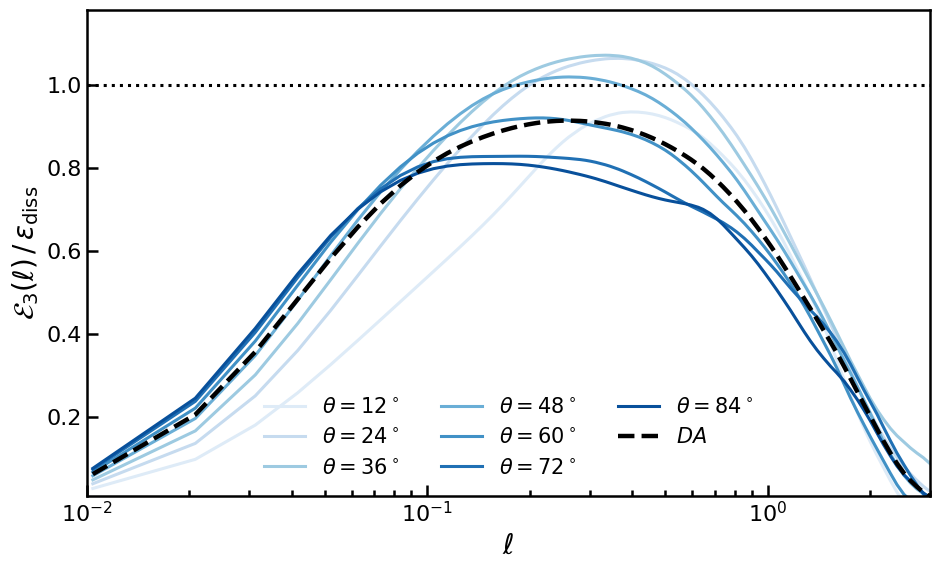}
  \caption{Scale-dependent third-order estimate $\mathcal{E}_{3}(\boldsymbol{\ell})$ (Eq.~\eqref{eq:E3_def}) for different lags, normalized by the true dissipation rate $\varepsilon_{\rm diss}=1.66$. Each colored curve corresponds to one polar-angle $\theta$, obtained by azimuthally averaging the six directions that share the same $\theta$ (i.e., $\phi$-average). The black dashed curve shows the DA result obtained by performing the solid-angle average over all sampled directions with the appropriate $\sin\theta$ weighting (Eq.~\eqref{eq:discre_Y}). The black dotted line at unity is plotted for reference.}
  \label{fig:theta_dependence}
\end{figure}

Recall that, we are collecting data along 42 directions (7 polar angles and 6 azimuthal angles), i.e., 42 independent trajectories. 
Based on Eq.~\eqref{eq:E3_def}, each sampled direction yields one $\mathcal{E}_{3}(\boldsymbol{\ell})$ curve. For each direction,
the reported curve is obtained by averaging the results over four virtual spacecraft.  To show $\theta$ dependence, we average the curves over the
six azimuths and obtain 7 curves at different $\theta$s, as shown in Figure~\ref{fig:theta_dependence}. One can see that for each curve, $\mathcal{E}_3$ is very small at the dissipation scale, increases to the peak at the inertial range, and decreases again at the energy containing range. The curves along different polar angles do not collapse with each other, which indicates a clear $\theta$ dependence. In particular, the peak level and position of the curves varies with polar angles. If we use the peak of the azimuthal-averaged curve as the cascade rate estimate, this estimate is sensitive to the sampling direction (i.e., polar angle $\theta$) relative to the mean magnetic field $\mathbf{B}_0$ and would be off by $\pm 20\%$ from the actual dissipation rate.

{To alleviate the uncertainty in the cascade rate arising from estimating it only using a single direction, the result from the DA method as implemented in Eq.~\eqref{eq:discre_Y} is also shown in Figure~\ref{fig:theta_dependence}. The peak of the DA curve is very close to (though not exactly equal to) the true dissipation rate from Eq.~\eqref{eq:eps_total_def}, $\textrm{peak of }\mathcal{E}_3/\varepsilon_{\rm diss}\simeq 0.92$. The inertial range can also be identified through the DA curve, say when $\mathcal{E}_3/\varepsilon_{\rm diss}>0.85$. The identified inertial range is $0.13<\ell<0.50$, which is roughly consistent with the inertial range identified from energy spectra in Fig.~\ref{fig:energy_spectrum}, through $ l\sim1/k$. We anticipate that for a simulation with a longer inertial range, the DA method could give rise to a more accurate estimate.}

One may also note that among the seven polar angles, the $\theta\approx60^\circ$ curve agrees very well with the DA curve.
Consistently, in Figure~\ref{fig:phi_dependence} (right), after averaging over azimuthal angles, the estimate of the cascade rate
at $\theta=60^\circ$ is the closest to the theoretical value. This suggests that, in
a situation where full solid-angle coverage is difficult to achieve (as in many real observations), sampling near this polar
angle may provide a practical compromise between directional coverage and estimator robustness. More details about the specialty of  $\theta\approx60^\circ$ are included in \cite{jiang2025angular}.

\subsection{Azimuthal Angle ($\phi$) dependence}
\begin{figure}[H]
  \centering
  \begin{minipage}{0.48\textwidth}
    \centering
    \includegraphics[width=\linewidth]{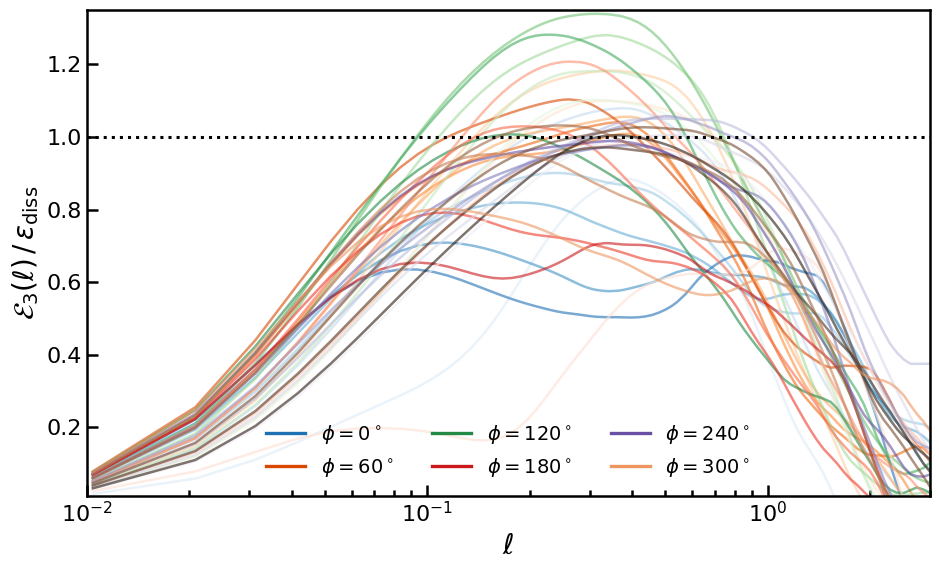}
  \end{minipage}
  \hfill
  \begin{minipage}{0.48\textwidth}
    \centering
    \includegraphics[width=\linewidth]{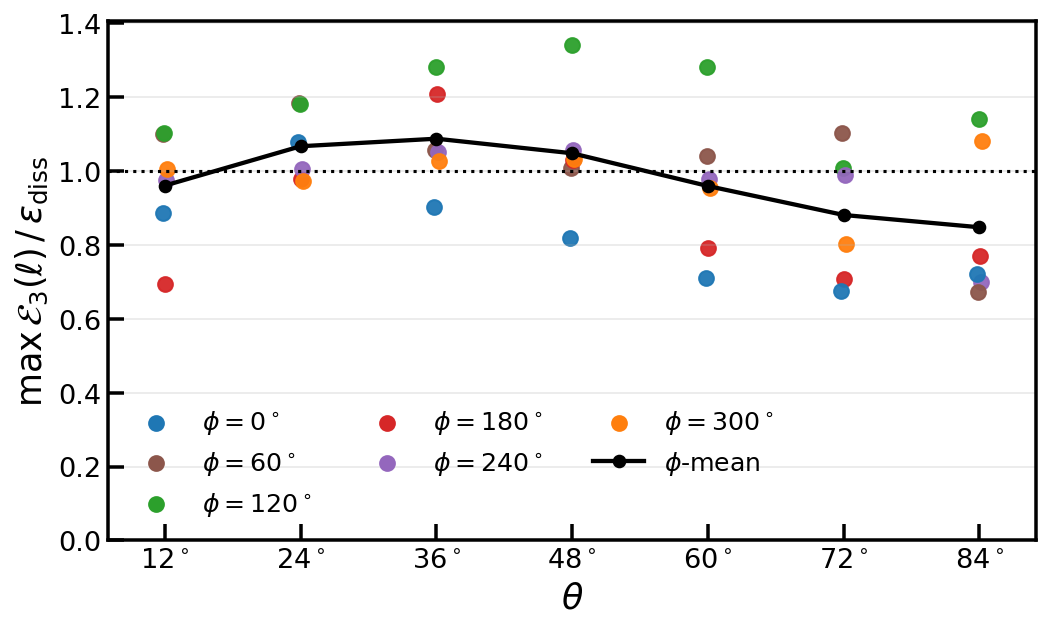}
  \end{minipage}
  \caption{Left: Scale-dependent third-order estimates $\mathcal{E}_{3}(\boldsymbol{\ell})$ (Eq.~\eqref{eq:E3_def}) for all 42 sampled directions, normalized with the true dissipation rate $\varepsilon_{\rm diss}=1.66$ and
  color-coded by azimuthal angle $\phi$; within each color there are seven curves corresponding to the seven sampled
  polar angles $\theta$. The horizontal black dashed line at unity is plotted for reference.
  Right: peak values of $\mathcal{E}_{3}(\boldsymbol{\ell})$ of curves on the left; each point corresponds to one direction and the
  black points indicate the azimuthal mean within each polar-angle group.}
  \label{fig:phi_dependence}
\end{figure}

In addition to the polar-angle dependence, we also observe a non-negligible azimuthal dependence. Figure~\ref{fig:phi_dependence} (left) shows the $\mathcal{E}_{3}(\boldsymbol{\ell})$ curves for all 42 directions. Curves sharing the same azimuthal angle $\phi$ are plotted in one color, so that each color with seven degrees of intensities indicates the seven sampled polar angles $\theta$. This visualization makes the $\phi$-dependence apparent: the curve shapes and their peak values differ systematically across azimuthal groups, indicating an intrinsic azimuthal anisotropy beyond the $\theta$ dependence. To quantify this trend, we take the peak value of each $\mathcal{E}_3(\boldsymbol{\ell})$ curve as the estimate of the cascade rate, and summarize the direction-by-direction values in Figure~\ref{fig:phi_dependence} (right). Each point corresponds to one direction and the black points show the azimuthal mean within each polar-angle group. For each polar angle, the azimuthal variation manifests as the vertical dispersion of the cascade rate estimate. When an adequate azimuthal averaging is performed, see the black points in Figure~\ref{fig:phi_dependence} (right), the polar angle dependence necessitates the appropriate polar-angle averaging, such as the DA method.

\subsection{Pros and cons of DA}
The energy transfer in the inertial range, quantified through the third-order law, shows observable angle dependence (anisotropy) in the presence of a mean magnetic field. The DA method can provide an accurate estimate of the energy dissipation rate through directional averaging of the third-order structure functions. 
The DA method shows weak dependence on the spacecraft separation $L_{\rm inter}$ in our setup.  This is expected because DA is constructed from along-trajectory increments at a prescribed lag and does not explicitly
rely on the inter-spacecraft baseline; with the four spacecraft following the trajectories with nearly the same direction (i.e., polar and azimuthal angles), changing the baseline
does not alter the sampled directions and therefore has only a minor effect on the feasibility and accuracy of the DA method. This has also been validated using the simulation data (not shown here), where the peak values of the $\phi$-averaged $\mathcal{E}_3(\boldsymbol{\ell})$ curves remain nearly unchanged across the six inter-spacecraft baseline families $L_{\rm inter}=\{10,\,20,\,30,\,40,\,50,\,60\}\,\ell_K$.
However, the DA method requires 3D information, i.e., the measurements at various $\theta$ and $\phi$. Consequently, when the
available angular coverage is limited, the DA method could become biased toward the sampled directions rather than
approaching the full solid-angle mean, and the estimated dissipation rate may deviate systematically from the true
value. In addition, DA relies on the Taylor hypothesis to convert temporal lags into spatial separations, which could introduce additional uncertainty.

\section{Results of Lag Polyhedral Derivative Ensemble (LPDE) Method}\label{sec:Results-LPDE}
After sampling all data points from the virtual spacecraft, we apply the LPDE procedure to the resulting four-spacecraft time series to estimate the lag-space divergence $-\frac{1}{4}\nabla_{\boldsymbol{\ell}}\!\cdot\boldsymbol{Y}^{\pm}$, as implemented in Eq.~\eqref{eq:eps_lpde}. For each sampling direction, we evaluate this quantity for every retained lag-tetrahedron, yielding one estimate of the cascade rate $\varepsilon_{LPDE}$ at the lag-tetrahedron mesocenter scale $\ell_{\mathrm{meso}}$. 
As discussed in Sec.\ref{sec:Methodology}, the LPDE method starts with the selection of tetrahedra in lag space, which suggests that the performance of LPDE could be significantly affected by the multispacecraft constellation. Therefore, we will investigate it in four aspects: inter-spacecraft separation $L_{\rm inter}$, quality of lag-tetrahedra (non-regularity level of lag-tetrahedra) $\sigma$, threshold $EP_{valid}$
($d_{EP}\le EP_{valid}$) used to retain lag-tetrahedra, and the trajectory of virtual spacecraft.

\subsection{Dependence on spacecraft separation}
To quantify how the inter-spacecraft separation influences LPDE, we evaluate LPDE performance across six baselines  $\{10,\,20,\,30,\,40,\,50,\,60\}\,\ell_K$. The inter-spacecraft separations of $L_{\rm inter}=\{\,50,\,60\}\,\ell_K$ are in the inertial range and $L_{\rm inter}=\{10,\,20,\,30,\,40\}\,\ell_K$ are roughly in the dissipation range. A fixed $\sigma=0.5$ and $EP_{valid}=0.85$ are used. We sample the simulation along 42 trajectories and produce 42 four-spacecraft time series realizations.

\begin{figure}[H]
  \centering
  \includegraphics[width=0.92\textwidth]{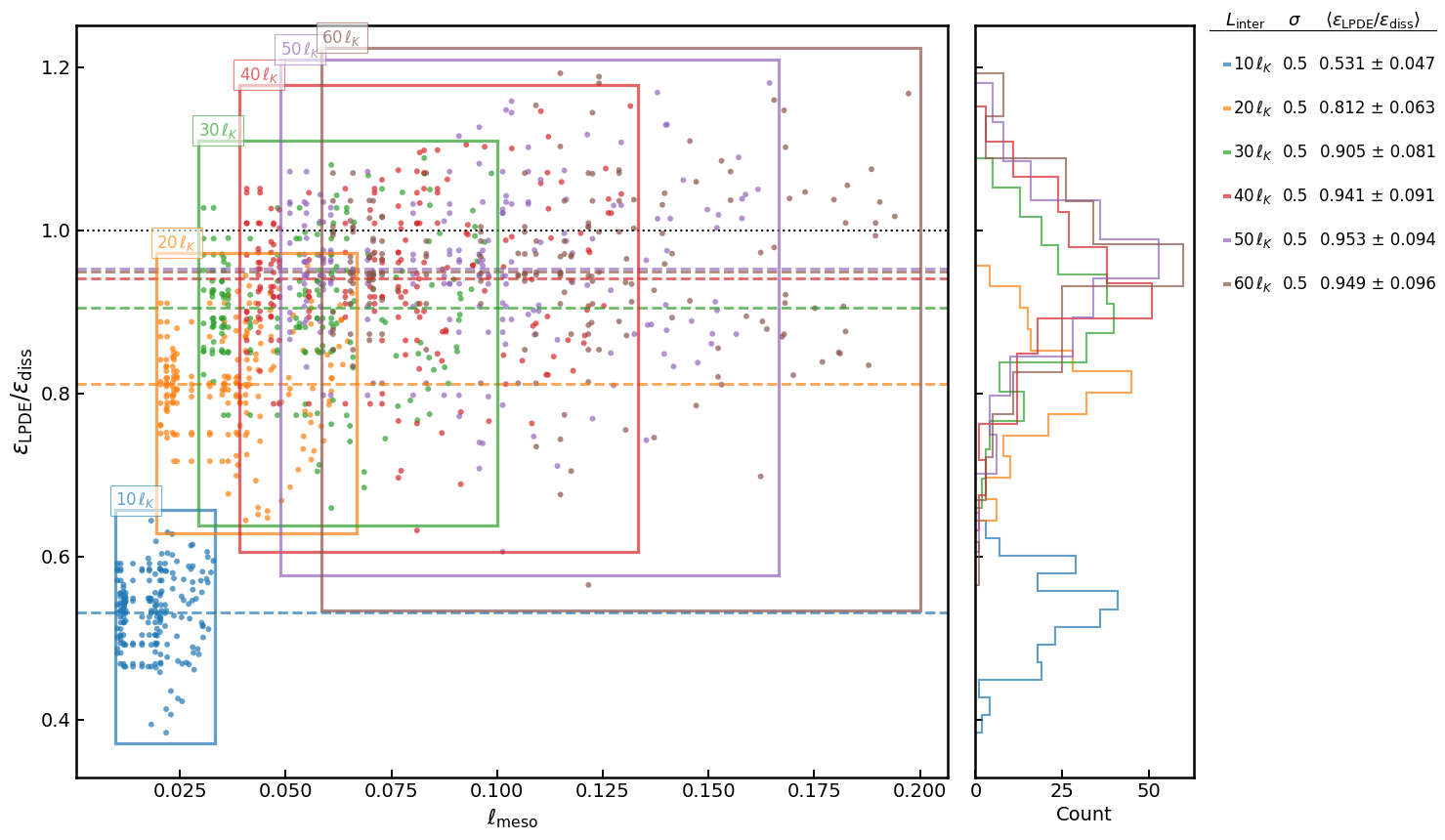}
  \caption{LPDE estimates of the cascade rate $\varepsilon_{LPDE}$ versus the lag-tetrahedron mesocenter length $\ell_{\mathrm{meso}}$ . The normalized cascade rate $\varepsilon_{LPDE}/\varepsilon_{\rm diss}$ is shown on the vertical axis.  Each dot represents one estimate from a retained lag-tetrahedron. Colors indicate different baseline families $L_{\rm inter}$. Also, colored boxes are used to outline the range of the point distribution for each baseline family. The horizontal dashed lines show the mean $\langle\varepsilon_{LPDE}/\varepsilon_{\rm diss}\rangle$ for each baseline family, while
the dotted black line marks $\varepsilon_{LPDE}/\varepsilon_{\rm diss}=1$. The right panel shows the corresponding histograms of
$\varepsilon_{LPDE}/\varepsilon_{\rm diss}$ for each baseline family.}
  \label{fig:comparison_scatter}
\end{figure}

Figure~\ref{fig:comparison_scatter} shows the LPDE estimates $\varepsilon_{LPDE}$ obtained from the ensemble of valid lag-tetrahedra, plotted as a
function of the lag-tetrahedron mesocenter scale $\ell_{\mathrm{meso}}$. The scatter shows that, for each baseline family, the mesocenters of lag-tetrahedra in lag space populate a band of scales. For example, when $L_{\rm inter}=50\ell_K$, the minimum $\ell_{\mathrm{meso}}\sim0.05$ and the maximum $\ell_{\mathrm{meso}}\sim 0.17$.  As we can see in Figure ~\ref{fig:comparison_scatter}, the mean $\varepsilon_{LPDE}$ varies systematically with the inter-spacecraft separation $L_{\rm inter}$, and
the agreement with the true dissipation rate improves as the spacecraft 
separation approaches the inertial range, see for example $L_{\rm inter}=\{\,50,\,60\}\,\ell_K$. This behavior is expected: changing the inter-spacecraft baseline changes the set of lag vectors available to LPDE and, therefore, the distribution of tetrahedron mesocenter scales in lag space. When the baseline is too small or too large, the resulting mesocenter scale distribution is shifted away from the inertial range (toward the dissipation range at small
baselines and toward the energy-containing scales at large baselines), and often becomes broader as well. As a result, the estimated cascade rate based on the third-order law that is valid in the inertial range is biased away from the true dissipation rate. 
The estimated cascade rate at $L_{\rm inter}=50\ell_K$ is about $\langle \varepsilon_{LPDE}/\varepsilon_{\rm diss}\rangle=0.95$. We anticipate that for a simulation with a longer inertial range, $\langle \varepsilon_{LPDE}/\varepsilon_{\rm diss}\rangle$ will get closer to 1.

\subsection{Offset (shape) effects}
\begin{figure}[H]
  \centering
  \includegraphics[width=0.7\textwidth]{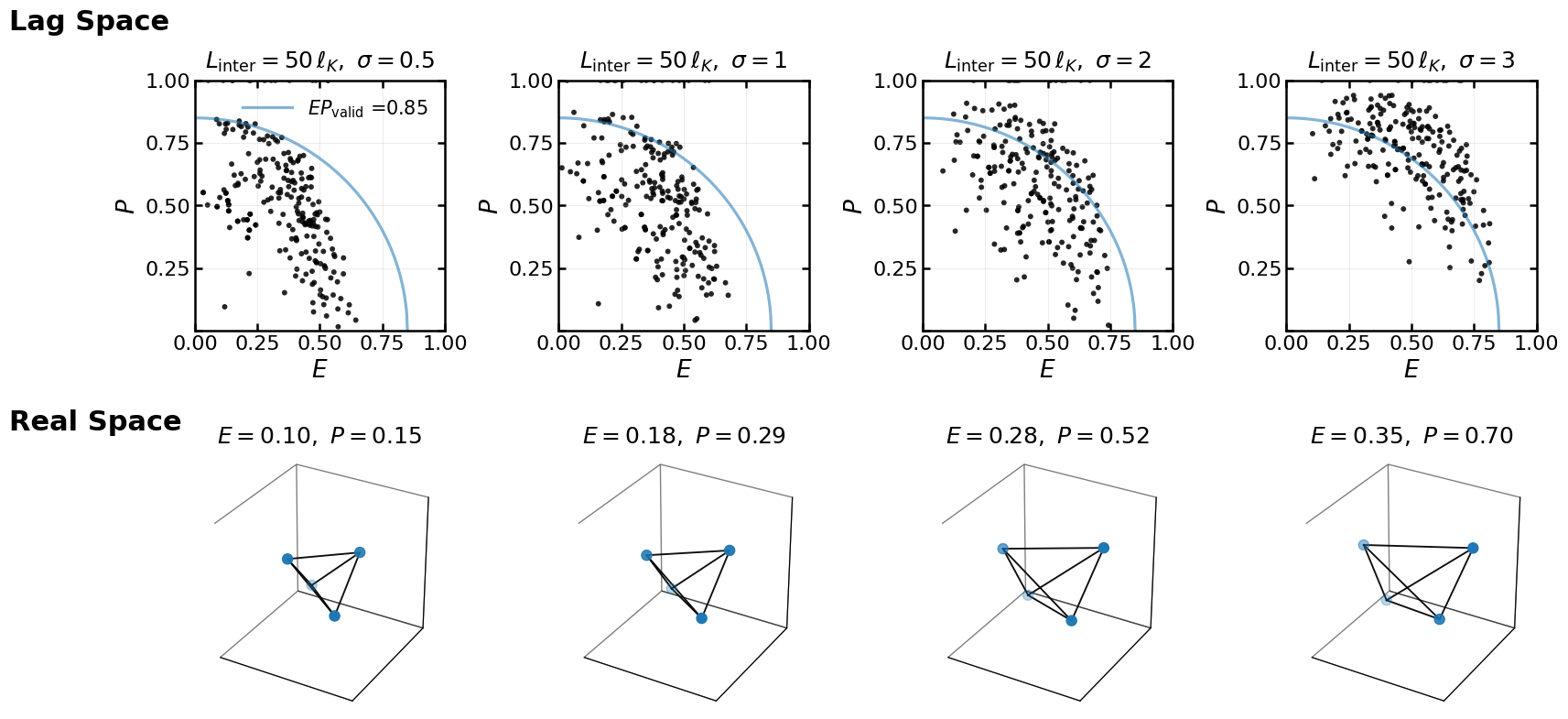}
  \caption{Top: $(E,P)$ of lag-tetrahedra with non-regularity levels $\sigma$ from $0.5$ to $3.0$. Each black dot represents one tetrahedron sampled in lag space and its $(E,P)$, and the blue curve marks the threshold $EP_{valid}=0.85$
($d_{EP}\le EP_{valid}$)  used to select valid lag-tetrahedra; Bottom: The corresponding spacecraft configurations in real space for different levels of non-regularity and the associated $(E,P)$ values are also listed.}
  \label{fig:scatter_comparison}
\end{figure}

To further explore the offset effect, we also test virtual spacecraft offsets forming irregular tetrahedra with different non-regularity levels $\sigma$. We fix $L_{\rm inter}=50\ell_K$ and $EP_{valid}=0.85$. Four non-regularity levels $\sigma=0.5, 1.0, 2.0, 3.0$ are used.
One can expect that the higher non-regularity levels will lead the tetrahedron in real space to become more irregular. Furthermore, this growing non-regularity also shows an impact on the lag-tetrahedra in lag space.  As shown in Figure~\ref{fig:scatter_comparison}, the planarity (P) and elongation (E) of the tetrahedron in lag space shift away from the low-$(E,P)$ region, indicating a reduced fraction of high-quality (well-conditioned) lag-tetrahedra.

\begin{figure}[H]
  \centering
  \includegraphics[width=0.95\textwidth]{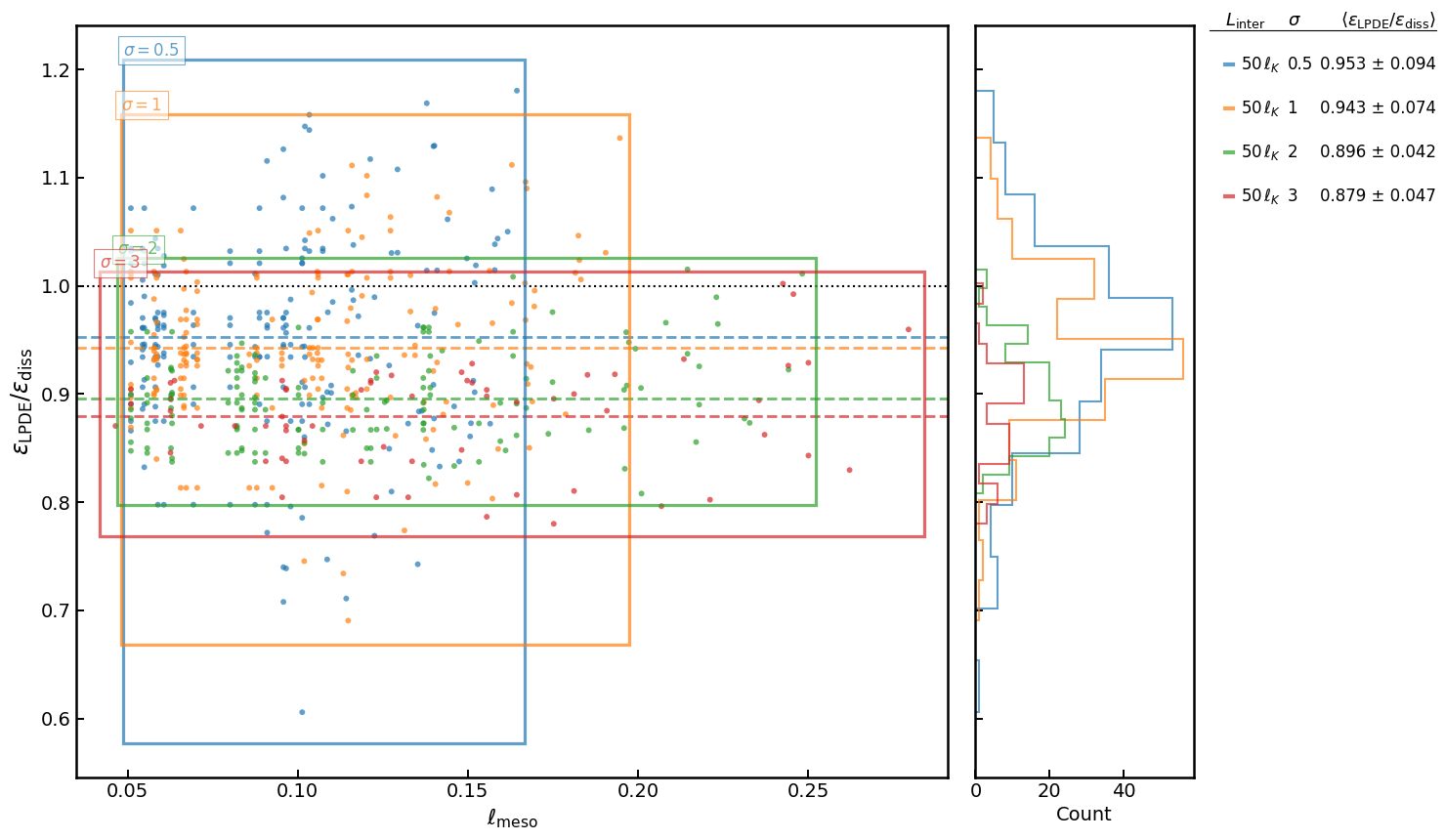}
  \caption{LPDE estimates of the cascade rate $\varepsilon_{LPDE}$ versus the lag-tetrahedron mesocenter length $\ell_{\mathrm{meso}}$ . The normalized cascade rate $\varepsilon_{LPDE}/\varepsilon_{\rm diss}$ is shown on the vertical axis.  Each dot represents one estimate from a retained lag-tetrahedron. Colors indicate different non-regularity level $\sigma$. Also, colored boxes are used to outline the range of the point distribution for each $\sigma$ case. The horizontal dashed lines show the mean $\langle\varepsilon_{LPDE}/\varepsilon_{\rm diss}\rangle$ for each $\sigma$ case, while
the dotted black line marks $\varepsilon_{LPDE}/\varepsilon_{\rm diss}=1$. The right panel shows the corresponding histograms of
$\varepsilon_{LPDE}/\varepsilon_{\rm diss}$ for each $\sigma$ case.}
  \label{fig:shape_comparison}
\end{figure}

Figure~\ref{fig:shape_comparison} shows that increasing non-regularity levels $\sigma$ can extend the mesocenter-length $\ell_{\mathrm{meso}}$ coverage (i.e., a broader
range of scales is sampled), over which the mean of the dissipation estimate is smaller than the true dissipation rate. Even though the mean for the $\sigma=0.5$ case is more accurate, the estimates are more vertically spread, thus a larger variance.
This highlights that LPDE is sensitive not only to the inter-spacecraft separation but also to tetrahedral quality.

\subsection{Trajectory independence}
\begin{figure}[H]
  \centering
  \includegraphics[width=0.48\textwidth]{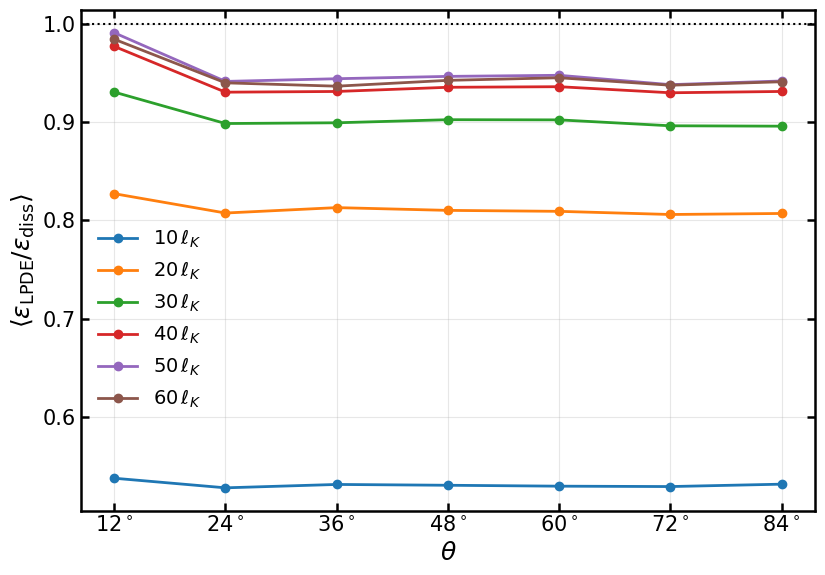}
  \caption{LPDE estimates of the cascade rate $\langle\varepsilon_{LPDE}\rangle$ versus the polar angle $\theta$ of virtual spacecraft trajectory. The normalized cascade rate $\langle\varepsilon_{LPDE}/\varepsilon_{\rm diss}\rangle$ is shown on the vertical axis. Colors indicate different baseline families $L_{\rm inter}$. Each dot represents azimuthally averaged normalized cascade rate within one polar angle group for a specific $L_{\rm inter}$. The dotted black line marks $\langle\varepsilon_{LPDE}/\varepsilon_{\rm diss}\rangle=1$. }
  \label{fig:trajectory_independence}
\end{figure}

As discussed in Sec.\ref{sec:virtual}, here we are flying the virtual spacecraft through the simulation domain along 42 directions arranged in seven polar angles
and six azimuthal angles. The LPDE estimate does not show a clear dependence on the trajectory direction in our experiments. We obtain the LPDE estimate from each trajectory, average over the six azimuthal angles (with equal weight for each azimuthal angle), and obtain the estimate for each polar angle.  As shown in
Figure~\ref{fig:trajectory_independence}, the LPDE estimates vary weakly with $\theta$. This is consistent with
the LPDE formulation: The estimator is based on the divergence of the Yaglom flux vector in lag space, where the lag vectors are set by the inter-spacecraft separations. As a scalar quantity, this divergence is independent of the sampling direction, provided that sufficiently ergodic coverage is attained.

\subsection{Threshold effect}
\begin{figure}[H]
  \centering
  \includegraphics[width=0.55\textwidth]{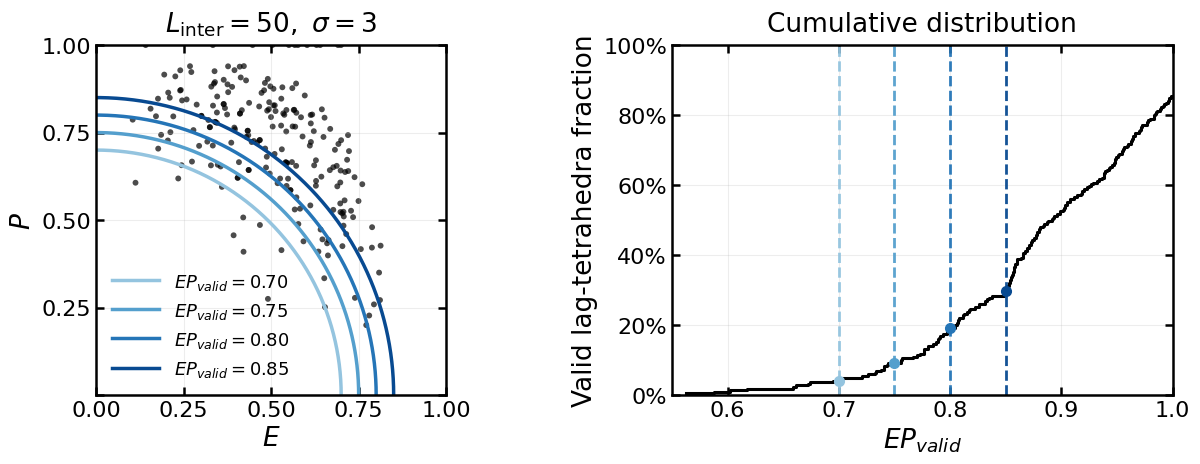}
  \caption{Left:$(E,P)$ of lag-tetrahedra for different thresholds $EP_{\rm valid}$. Each black dot represents one lag-tetrahedron sampled in lag space and its $(E,P)$, and the blue curve marks the threshold $EP_{\rm valid}$. Right: The cumulative distribution of the number of valid lag-tetrahedra as a function of $EP_{\rm valid}$. Vertical dashed lines represent four values of $EP_{\rm valid}=0.70, 0.75, 0.80, 0.85$.}
  \label{fig:threshold}
\end{figure}

One of the key steps to implement LPDE is the selection of lag-tetrahedra, as discussed in Sec.~\ref{sec:LPDE}. Only lag-tetrahedra satisfying $d_{EP}\le EP_{\rm valid}$ will be retained, which
removes highly elongated or nearly planar lag-tetrahedra while keeping a well-conditioned
ensemble for the subsequent calculation. 
To determine the role that the threshold $EP_{\rm valid}$ plays in the LPDE procedure, we also test different thresholds $EP_{\rm valid}=0.70, 0.75, 0.80, 0.85$ with fixed $L_{\rm inter}=50\ell_K$ and $\sigma=3$. As shown in Figure ~\ref{fig:threshold}, with larger $EP_{\rm valid}$, more lag-tetrahedra satisfy $d_{EP}\le EP_{\rm valid}$ and contribute to the estimation of cascade rate. From Table~\ref{tab:tetra_by_thr}, we could see that there is no significant variance of the estimated cascade rate. 

\begin{table}[H]
\centering
\begin{tabular}{c c c c}
\hline
Threshold $EP_{\rm valid}$ & Valid lag-tetrahedra & Used lag-tetrahedra &$\langle\varepsilon_{LPDE}/\varepsilon_{\rm diss}\rangle$\\
\hline
0.70 & 9  & 8  & $0.879 \pm 0.047$ \\
0.75 & 20 & 19 & $0.869 \pm 0.047$ \\
0.80 & 42 & 40 & $0.873 \pm 0.045$ \\
0.85 & 65 & 63 & $0.875 \pm 0.045$ \\
\hline
\end{tabular}
\caption{Lag-tetrahedra statistics for four $EP_{valid}$ thresholds and the corresponding estimated cascade rate $\langle\varepsilon_{LPDE}/\varepsilon_{\rm diss}\rangle$. Valid lag-tetrahedra column presents how many lag-tetrahedra satisfy the selection criteria $d_{EP}\le EP_{\rm valid}$; Used lag-tetrahedra column shows how many lag-tetrahedra satisfying the selection criteria have non-zero mesocenter lengths.}
\label{tab:tetra_by_thr}
\end{table}

\subsection{Pros and cons of LPDE}
In summary, the LPDE method takes into account the anisotropic effect on the third-order law, which therefore can provide a reliable estimate of the energy cascade rate, given that the third-order law is valid in the inertial range. In contrast to the DA method, two practical advantages of LPDE are that: (i) it does not require directional coverage in real space (since it computes the
divergence of the Yaglom flux vector in lag space), therefore it does not show spacecraft trajectory dependence. (ii) It does not rely on the Taylor hypothesis because the lags in the calculation are from the inter-spacecraft separations.
However, the LPDE estimate of the cascade rate is sensitive to the inter-spacecraft separation and to the tetrahedral quality. The  LPDE estimate of the cascade rate is based on the validity of the third-order law, which holds only within (an approximate) inertial range. When the spacecraft baselines drift toward the energy-containing scales or the dissipation range, additional terms in the vKH (Eq.\ref{eq:VKH_MHD})  are non-negligible, and the third-order law gradually breaks down. Consequently, LPDE estimates obtained from such baselines may show systematic departures from an inertial-range plateau even when the divergence itself is accurately computed. Such departures do not necessarily imply a deficiency of the LPDE method; instead, they reflect the limited applicability of the third-order law outside the inertial range, which only captures the nonlinear contribution to the cascade rate.



\section{Conclusion and Discussion}\label{sec:Conclusion and Discussion}
We made a systematic comparison between two estimators of the energy cascade/dissipation rate in incompressible MHD turbulence: the direction-averaging (DA) method and the lag polyhedral derivative ensemble
(LPDE) method. Using a three-dimensional driven MHD simulation with a mean magnetic field $\mathbf{B}_0\,\hat{\mathbf{z}} = 2$,
we constructed virtual four-spacecraft time series and investigated in detail the accuracy of the two estimators of the energy cascade rate.

\begin{figure}[H]
  \centering
  \includegraphics[width=0.48\textwidth]{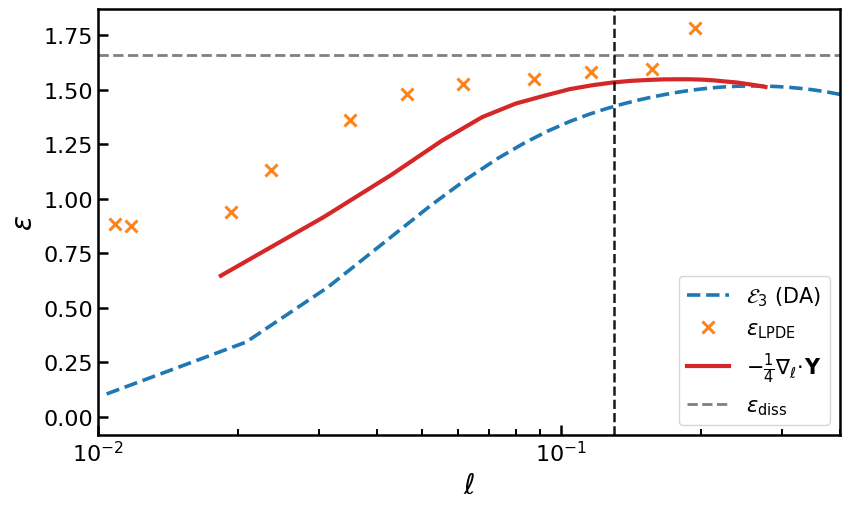}
  \caption{Scale-dependent energy cascade rate from the third-order law using different methods. The red solid line, serving as the reference curve, is the grid-computed $-\frac{1}{4}\nabla_{\boldsymbol{\ell}}\!\cdot\mathbf{Y}$. The orange crosses show the LPDE estimate, which is the average of points in Fig.~\ref{fig:comparison_scatter} over scale bins. The blue dashed curve shows the result of 
  DA, which is also shown in Figure~\ref{fig:theta_dependence}. The horizontal dashed line indicates the true dissipation rate $\varepsilon_{\rm diss}=1.66$. The vertical dashed line indicates the begin of inertial range identified by DA method.}
  \label{fig:_grid_comparison}
\end{figure}

In principle, both DA and LPDE can provide reliable estimates of the energy cascade rate in anisotropic MHD turbulence. The benchmark for evaluating the accuracy of these methods is the reference red curve in Figure~\ref{fig:_grid_comparison}, which is the direct computation of $-\frac{1}{4}\nabla_{\boldsymbol{\ell}}\!\cdot\mathbf{Y}^{\pm}$ using the simulation fields at the grid points via a second-order central finite difference scheme. Also shown in Figure~\ref{fig:_grid_comparison} are the curves from the LPDE with $\sigma=0.5$ and $EP_{\rm valid}=0.85$ and DA methods. 
All three curves show consistent estimation toward each other as the lag enters the inertial-range scales. Small deviations from the reference curve could arise from several sources. For the LPDE, each point is associated with the lag-tetrahedron mesocenter, which is the arithmetic mean of four lag vectors. For the DA, here we only use 42 uniformly distributed directions, which could be refined by incorporating information from more directions.

Our results clarify the complementary strengths and limitations of the two estimators. The DA method is limited by angular coverage: when the lag directions do not adequately sample the full
$4\pi$ solid angle, the estimated cascade rate could deviate significantly from the true dissipation rate. In Figure~\ref{fig:phi_dependence}, the longitudinal third-order structure function curves vary significantly with $\theta$ and $\phi$, demonstrating directional dependence of energy transfer in anisotropic turbulence. The DA method averages over the contributions from a number of directions, which alleviates the bias of the estimate along a single direction.
In contrast, the LPDE method exhibits weak dependence on the sampling direction ($\theta$ or $\phi$) in our experiments, and instead is primarily limited by the spacecraft constellation: the inter-spacecraft separation should be close to the inertial range, and the tetrahedra must remain sufficiently well-conditioned. Since the estimator ultimately relies on the third-order law in the inertial range, when the inter-spacecraft separation falls outside the inertial range, the LPDE leads to biased dissipation estimates. 

Possible refinements can further improve both estimators. For the DA method, the estimate can be significantly improved by better angular coverage, which has been implemented using MMS and Cluster data, see for example, \cite{Osman11-PRL,BandyopadhyayEA18-msheath}.
In addition, as suggested by the results in \citet{jiang2025angular}, among different $\theta$ directions, the estimate along $\theta\simeq 60^\circ$ is more accurate than other directions. This suggests that future DA implementations could benefit from non-uniform weighting schemes, in which directions with greater physical relevance are assigned larger weights instead of treating all sampled directions equally. For the LPDE method, possible improvements include a more selective treatment of the lag-tetrahedra. In addition to simple geometric thresholds we are using here, one may adopt more sophisticated strategies that account simultaneously for tetrahedral quality and characteristic size in lag space \citep{Broeren2021}. Different lag-tetrahedra may also be assigned different weights, rather than being combined uniformly, so that better-conditioned lag-tetrahedra or lag-tetrahedra with characteristic scale in the inertial range contribute more to the final estimate. 

Our findings have direct implications for current and future multi-spacecraft missions. Constellations such as MMS, HelioSwarm, and Plasma Observatory provide a natural pathway to realize the LPDE estimate, but reliable results depend on baseline planning that targets inertial-range separations and maintains adequate, well-conditioned tetrahedra. At the same time,  the DA estimate remains valuable, especially when angular coverage can be accumulated over time or across multiple trajectories.

\begin{acknowledgments}
    This work was supported by the NASA award 80NSSC25K7757 and the University of Delaware General University Research Program grant.
    F.P. is partially supported by NSF SHINE 2501387 at the University of Delaware.
    J.E.S. is supported by the Royal Society University Research Fellowship awards URF\textbackslash R\textbackslash 251029 and URF\textbackslash R1\textbackslash201286.
    We would like to acknowledge high-performance computing support from Cheyenne (https://doi.org/10.5065/D6RX99HX) and Derecho (https://doi.org/10.5065/qx9a-pg09) provided by NCAR's Computational and Information Systems Laboratory, sponsored by the National Science Foundation.
\end{acknowledgments}

\bibliographystyle{aasjournalv7}
\bibliography{A-Z,refs_YY}


\end{CJK*}
\end{document}